\documentclass[aps,preprint,nofootinbib,superscriptaddress]{revtex4}

\usepackage{mathrsfs}
\usepackage{amsmath}
\usepackage{array}
\usepackage{verbatim}
\usepackage{epsfig}
\usepackage{epsf}
\usepackage{CJK}
\usepackage{graphicx}

\begin{document}

\title{ Coulomb energy of axially deformed nucleus}

\author{Ning Wang}
\affiliation{Department of Physics, Guangxi Normal University,
Guilin 541004, P. R. China}

\author{Xuexin Yu}
\affiliation{Department of Physics, Guangxi Normal University,
Guilin 541004, P. R. China}

\author{Min Liu\footnote{Corresponding author : lium$\_$816@hotmail.com}}
 \affiliation{Department of Physics,
Guangxi Normal University, Guilin 541004, P. R. China}
\affiliation{ College of Nuclear Science and Technology, Beijing
Normal University, Beijing, 100875, P. R. China}

\begin{abstract}
We previously proposed a formula for calculating the Coulomb
energy of spherical nucleus with Wood-Saxon charge distribution.
In this work, the analytical formula is extended for description
of the Coulomb energy of nucleus with $\beta_2$ deformation.
\end{abstract}

\maketitle

The calculation of the Coulomb energy for complicated charged
system with small computing effort and high accuracy is a great
challenge in physics and quantum chemistry research
\cite{Yu,D98,Man}. For a system with an arbitrary charge
distribution $\rho({\bf r})$, the direct term of the Coulomb
energy can be calculated with
\begin{eqnarray}
E_{C}=\frac{e^{2}}{2}\int\!\!\int\frac{\rho({\bf r})\rho({\bf
r'})} {|{\bf r}-{\bf r'}|}d {\bf r}d {\bf r'} .
\end{eqnarray}
However, the six-dimensional integration in Eq.(1) is very
time-consuming and becomes a bottleneck in the large-scale
calculations of potential energy surfaces of nuclear systems. In
this work, we attempt to propose an analytic expression for
calculating the Coulomb energy of nucleus with both the nuclear
surface diffuseness and nuclear $\beta_2$ deformation being taken
into account. For reader's convenience, the approach to calculate
the Coulomb energy of a nuclear system proposed in our previosuly
work \cite{Yu} is reviewed firstly, and then the analytical
formula for calculating the Coulomb energy of spherical nucleus
with Woods-Saxon density distribution will be extended for
description of the Coulomb energy of nucleus with $\beta_2$
deformation.

The Coulomb energy of an arbitrary nuclear system can be obtained
by
\begin{eqnarray}
E_{C}=\frac{e}{2}\int\rho({\bf r})V_{C}({\bf r})d{\bf r},
\end{eqnarray}
where $V_{C}({\bf r})$ is the Coulomb potential which is obtained
by solving the Poisson equation
\begin{eqnarray}
\nabla^{2}V_{C}({\bf r})=-4\pi e\rho({\bf r}).
\end{eqnarray}
The charge distribution of a nucleus is usually described by a
Woods-Saxon form,
\begin{eqnarray}
\rho(r)=\frac{\rho_{0}}{1+\exp  ( \frac{r- \mathcal {R}} {a}
  )}.
\end{eqnarray}
Where, $\rho_{0}$ and $a$ denote the central charge density and
the surface diffuseness, respectively. $\mathcal {R}$ defines the
distance from the origin of the coordinate system to the point on
the nuclear surface. For an axially deformed system, $\mathcal
{R}$ is expressed as,
\begin{eqnarray}
\mathcal {R} (\theta)= R_0 \, [1+  \beta_{2} Y_{20}(\theta) +
...].
\end{eqnarray}
In the calculation of Coulomb energy of nucleus as a function of
nuclear deformation, we remain the central charge density
$\rho_{0}$ of the nucleus unchanged by using the conservation of
charge number and varying the half-density radius $R_0$ to
consider the effect of incompressibility of nuclear matter in the
nucleus.

The Poisson equation is solved by a code \textbf{hwscyl} (a
Fortran subroutine in FISHPACK \cite{Paul99}) which is an adaptive
fast solver for solving a five-point finite difference
approximation to the modified Helmholtz equation in cylindrical
coordinates using a centered finite difference grid. We calculate
the Coulomb potential in cylindrical coordinates within a region
$x=0\sim40$ fm and $z=-40\sim40$ fm (using a grid with step size
$0.1$ fm). It is known that when $r \gg \mathcal {R}$, the
asymptotic behavior of the Coulomb potential of a nucleus is
$V_C=eZ/r$, which gives the boundary condition in solving the
Poisson equation.  The Coulomb energy of an arbitrary axially
deformed nuclear system can be obtained with a  two-dimensional
integration over the Coulomb potential $V_{C}({\bf r})$ which can
be calculated with the very fast solver for the Poisson equation
mentioned above.

In our previous work \cite{Yu}, we investigated the Coulomb
energies of spherical nuclei with Wood-Saxon charge distributions.
The central charge density of a nucleus is obtained with the
Skyrme energy density functional together with the extended
Thomas-Fermi (ETF) approach \cite{Liu06}. The nuclear surface
diffuseness $a$ varies from $0.1 $ to $1.2 $ fm in which the
central charge density is remained unchanged. We find that the
Coulomb energies of spherical nuclei with Woods-Saxon charge
distributions can be well described with an analytical expression
based on the leptodermous expansion \cite{RAI88},
\begin{eqnarray}
E_{\rm Coul}=E^{(0)}_{C}F (\omega)
\end{eqnarray}
with \cite{Yu}
\begin{eqnarray} F (\omega) = 1-\frac{5}{2}\omega^{2}+c_3
\omega^{3}+\omega^{4}+ c_5 \omega^{5} + c_6 \omega^{6}+...
\end{eqnarray}
Where, $\omega=\frac{\pi}{\sqrt 3}\frac{a}{R}$ and $E^{(0)}_{C}$
denotes the Coulomb energy of  a spherical nucleus with uniform
charge distribution,
\begin{eqnarray}
E^{(0)}_{C}=\frac{3}{5}\frac{Z^{2}e^{2}}{R}.
\end{eqnarray}
$Z$ denotes the charge number of the nucleus and $R=\left[
Z/(\frac{4 \pi}{3} \rho_0) \right ]^{1/3}$ is the corresponding
radius of a spherical nucleus with uniform charge distribution. By
fitting the calculated Coulomb energies with numerical
integrations for a number of spherical nuclei along the
$\beta$-stability line, we obtained the coefficients $c_3=3.005$,
$c_5=-4.822$, $c_6=2.934$.

\begin{figure}
\includegraphics[angle=-0,width=1.0\textwidth]{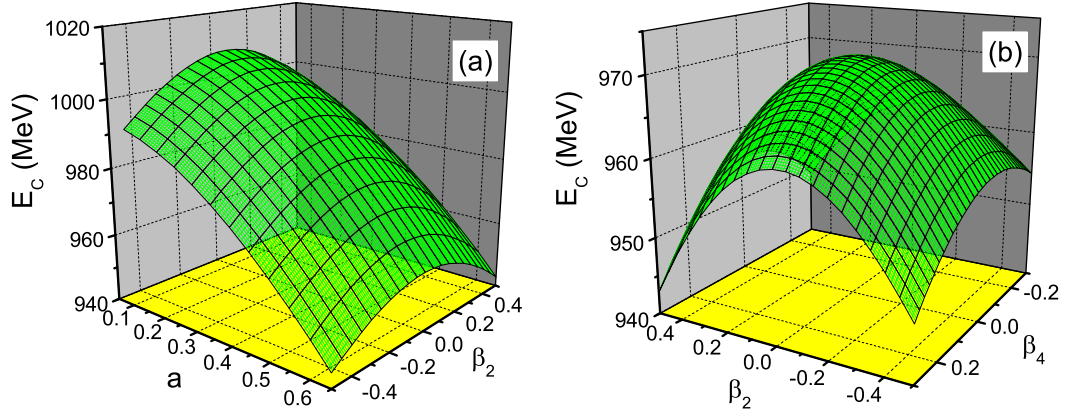}
\caption{(Color online) Coulomb energies of $^{238}$U as a
function of  nuclear surface diffuseness $a$ and quadrupole
deformation $\beta_2$ (a); and as a function of  $\beta_2$ and
$\beta_4$ (b). }
\end{figure}

\begin{figure}
\includegraphics[angle=-0,width=1.0\textwidth]{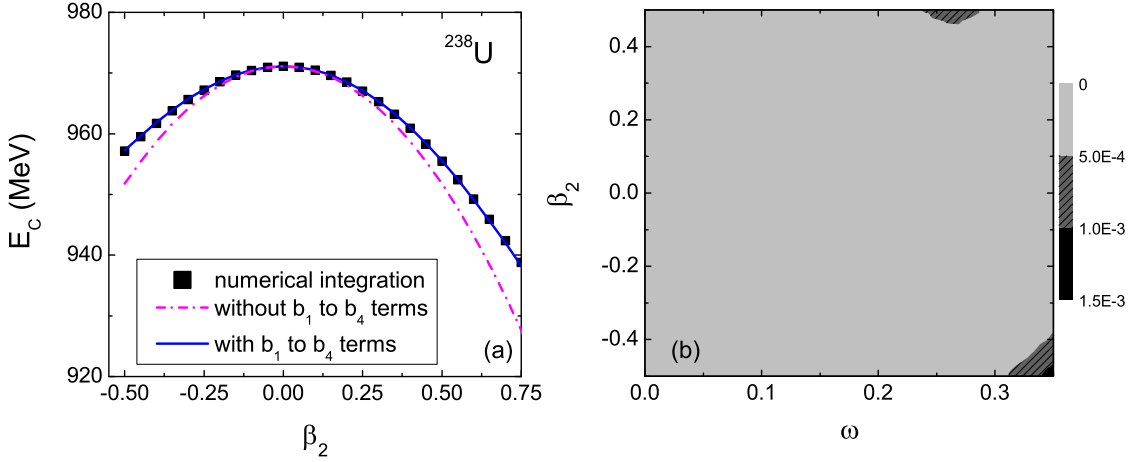}
\caption{(Color online) (a) Coulomb energy of $^{238}$U as a
function of $\beta_2$ deformation (with $a=0.55$ fm). The squares
denote the results with numerical integration [with Eq.(2)]. The
dot-dashed curve and the solid curve denote the results of Eq.(9)
without and with the $b_1$ to $b_4$ terms being taken into
account, respectively. (b) Relative deviations $|E_{\rm
Coul}-E_C|/E_C$ of the Coulomb energies from the numerical
integration results. $E_{\rm Coul}$ denotes the calculated Coulomb
energy with Eq.(9) and (10). The region with light gray denote
that the relative deviations are smaller than $0.05\%$.}
\end{figure}

With the same approach, we investigate the Coulomb energies of
deformed nuclei. In Fig.1 (a), we show the calculated Coulomb
energy of $^{238}$U as a function of nuclear surface diffuseness
$a$ and quadrupole deformation $\beta_2$, and we show the
corresponding Coulomb energy as a function of $\beta_2$ and
$\beta_4$ (with $a=0.55$ fm) in Fig.1(b). One can see that the
Coulomb energy decreases with increase of the nuclear surface
diffuseness and of the deformation. In this work, we write the
Coulomb energy of a nucleus as
\begin{eqnarray}
E_{\rm Coul}=E^{(0)}_{C}F (\omega) G (\omega,\beta),
\end{eqnarray}
with a factor $G (\omega,\beta)$ to consider the influence of
nuclear deformation. For nucleus with $\beta_2$ deformation, we
assume that the factor $G (\omega, \beta)$ has a form
\begin{eqnarray}
G (\omega, \beta_2) = 1 - \frac{1}{4 \pi} \beta_2^{2}  + b_1
\omega \beta_2^{2}+b_2 \omega^{2} \beta_2^{2}+ b_3 \beta_2^{3} +
b_4 \beta_2^{4}+...
\end{eqnarray}
Where the term $-\frac{1}{4 \pi}\beta_2^{2}$ is presented by
Greiner and Maruhn in Ref. \cite{Greiner}. By fitting the
calculated Coulomb energies with Eq.(2) as a function of surface
diffuseness ($a\le 0.7$ fm) and quadrupole deformation
($|\beta_2|\le 0.5$) for a number of nuclei along the
$\beta$-stability line, we obtain the coefficients $b_1=\frac{1}{4
\pi}$, $b_2= 0.188$, $b_3=-0.007$ and $b_4=0.018$.

\begin{table}
\caption{ Values of $\omega=\frac{\pi}{\sqrt 3}\frac{a}{R}$ for
some nuclei by taking $a=0.55$ fm and $R=1.2A^{1/3}$ fm. }
\begin{tabular}{cccccccc}
 \hline\hline
 &~~$^{16}$O ~~ & ~~$^{40}$Ca~~ & ~~$^{90}$Zr~~  & ~~$^{144}$Sm~~ & ~~$^{208}$Pb~~ & ~~$^{238}$U~~ & ~~$^{298}$114~~ \\
\hline
 ~~$\omega$~~ & 0.33    &   0.24     & 0.19     &  0.16      & 0.14         &  0.13     &  0.12  \\
\hline\hline
\end{tabular}
\end{table}

In Fig.2(a), we show the Coulomb energy of $^{238}$U as a function
of nuclear quadrupole deformation $\beta_2$. The squares denote
the results with Eq.(2). The dot-dashed curve and the solid curve
denote the results of Eq.(9) without and with the $b_1$ to $b_4$
terms being taken into account, respectively. One can see that the
higher-order terms of deformation are still required  for system
with large deformation. In Fig.2(b), we show the relative
deviations $|E_{\rm Coul}-E_C|/E_C$ of the Coulomb energies from
the numerical results for a number of nuclei $A=16 \sim 300$
varying the surface diffuseness ($a\le 0.7$ fm) and the quadrupole
deformation ($|\beta_2|\le 0.5$). $E_{\rm Coul}$ denotes the
calculated Coulomb energy with Eq.(9) and (10). From Fig.2, one
can see that the Coulomb energy obtained with the analytical
formula Eq.(9) is close to the calculated results with numerical
integration for most cases. In Table 1, we list some typical
$\omega$ values of a series of nuclei from light to heavy. For
intermediate and heavy nuclei, $\omega$ has a value about $0.1
\sim 0.25$, the corresponding relative deviations of the Coulomb
energies with Eq.(9) are smaller than $0.05\%$ [denoted by light
gray in Fig.2(b)] for almost all cases with $a\le 0.7$ fm and
$|\beta_2|\le 0.5$. For light nuclei, the corresponding values of
$\omega$ are larger than 0.25 in general and the relative
deviations of the Coulomb energies with Eq.(9) slightly increase
for some cases with strong deformations. It is known that the
charge distributions of light nuclei are usually described by
gaussian functions rather than the Woods-Saxon form. The
analytical expression of the Coulomb energy of a system with
gaussian charge distribution can be found in Refs.
\cite{RAI88,Yu}.

In summary, the Coulomb energy of axially deformed nucleus with
Wood-Saxon charge distribution has been investigated. The Coulomb
energy of a nuclear system was numerically calculated with a
two-dimensional integration over the Coulomb potential which was
obtained by solving the Poisson equation. By fitting the
numerically calculated Coulomb energies for a number of nuclei
from $A=16$ to 300 with $\beta_2$ deformed Wood-Saxon charge
distribution, an analytical formula, that is a function of nuclear
$\beta_2$ deformation and surface diffuseness, is finally
obtained. The relative deviation of the Coulomb energy with the
proposed formula is generally smaller than $0.05\%$ for
intermediate and heavy nuclei at normal deformations. For nuclear
system with $|\beta_2|>0.5$, such as fissioning system, the
proposed formula could not be applicable, and the two-dimensional
numerical integration over the Coulomb potential has to be
performed to obtain accurate results.

\begin{center}
\textbf{ACKNOWLEDGEMENTS}
\end{center}
This work is supported by National Natural Science Foundation of
China, Nos 10875031, 10847004. The code to calculate the Coulomb
energy of axially deformed system with numerical integrations over
the Coulomb potential is available from http://www.imqmd.com

\end{document}